\documentclass[a4paper,fleqn,usenatbib]{mnras}

\usepackage[T1]{fontenc}
\usepackage{ae,aecompl}

\usepackage{graphicx}	
\usepackage{multirow, array} 
\usepackage{float} 
\usepackage{placeins}
\usepackage{color}
\usepackage{subfig}
\usepackage{amssymb, amsmath, amsbsy} 

\title{ Determination of the superoutburst cycle lengths of 206 SU UMa type dwarf novae}

\author[Vogt et al.]{
N. Vogt,$^{1}$\thanks{E-mail: nikolaus.vogt@uv.cl (NV)}
E. C. Puebla,$^{1}$
and A. Contreras-Quijada$^{1}$\\
\\
$^{1}$Instituto de F\'isica y Astronom\'ia, Universidad de Valpara\'iso, Avda. Gran Breta\~na 1111, Valpara\'iso, Chile\\
}

\date{Accepted XXX. Received YYY; in original form ZZZ}

\pubyear{2020}

\begin{document}
\label{firstpage}
\pagerange{\pageref{firstpage}--\pageref{lastpage}}
\maketitle

\begin{abstract}
SU UMa stars are characterized by ``superoutbursts" which are brighter at maximum light and which last much longer than the more frequent ``ordinary" outbursts of these dwarf novae. Although there are now more than 1180 SU UMa type dwarf novae catalogued, our knowledge on their superoutburst cycle length C$_\mathrm{so}$ was hitherto limited to about 6$\%$ of the entire sample of known SU UMa stars. Using public data bases  we have determined new C$_\mathrm{so}$ values for a total of 206 additional SU UMa stars in the range 17 d < C$_\mathrm{so}$ < 4590 d (including some ER UMa and WZ Sge type representants) within total time intervals between  2  and 57 years, and with an estimated uncertainty of $\pm$11$\%$. This way, we are increasing our present knowledge of C$_\mathrm{so}$ values by a factor $\sim$3.8.  Its distribution is characterized by a broad maximum around C$_\mathrm{so}$ $\approx$ 270 days, and slowly decreasing numbers till C$_\mathrm{so}$ $\approx$ 800 d.  The domain C$_\mathrm{so}$ > 450 d was unexplored until now; we add here  106 cases ($\sim$51$\%$ of our total sample) in this range of long cycles, implying a better statistical basis for future studies of their distribution. Our sample contains 16 known WZ Sge stars, and we propose WZ Sge membership for 5 others hitherto classified as ordinary SU UMa stars. Individual superoutburst timings deviate in average about $\pm$7$\%$  of the cycle length from their overall linear ephemeris, conﬁrming a pronounced quasi-periodic repeatability of superoutbursts.  All relevant parameters are listed with their errors, and a table with individual superoutburst epochs of our targets is given, enabling future researchers to combine our results with other (past or future) observations.

\end{abstract}

\begin{keywords}
stars: novae, dwarf novae, cataclysmic variables.
\end{keywords}



\section{Introduction}

Dwarf novae (DNe) are cataclysmic variables (CVs) , i.e. interacting binaries, whose low-mass main sequence secondary is filling its Roche lobe, causing a permanent mass flow towards the white dwarf primary. Conservation of angular momentum of the transferred mass leads to the formation of an accretion disk. Instabilities within this disk are the principal cause of DN outbursts \citep{warner9}.  Among the DNe the so-called SU UMa sub-class is characterized by two very distinct outbursts types: short ones (lasting a few days) and superoutbursts which can last two weeks or longer in their rather bright ``plateau" phase. Typical superoutburst cycle lengths of these ``ordinary" SU UMa stars range from 100 to 500 days. Two additional sub-groups can be distinguished: WZ Sge stars have longer superoutburst cycle lengths (several thousand days)  and lack the normal, short outbursts, while ER UMa type dwarf novae are characterized by very short superoutburst cycles (20-40 days), and even more frequent normal outbursts in between. 

More than four decades ago, periodic ``superhumps" have been detected in the SU UMa star VW Hyi \citep{vogt1, warner2}  during a plateau phase; later they turned out to be a  characteristic feature of all SU UMa stars, but they appear occasionally also in other CV types, as classical nova remnants and nova-like stars \citep{patterson1}. Since then, and till now most observational efforts on SU UMa stars have been concentrated to detect superhumps and to determine their properties \citep[][ and previous papers of a series]{kato1}, widely  neglecting a systematic coverage of the general outburst behavior of these and other dwarf novae.

The semi-regular outburst behavior of DNe has been an appealing item of studies by amateur astronomers for a long time, organized by their societies, as the American Association of Variable Star Observers (AAVSO), the British Astronomical Association  (BAA)  and the Royal Astronomical Society of New Zealand, Variable Star Section (RASNZ,VSS), and very valuable long-term light curve studies have already been published decades ago, containing  catalogues of individual outbursts  with their  main properties as maximum brightness, amplitude,  width and classification according to light curve shape, as well as  some  statistical analysis of these properties.  As typical examples, we can mention here extensive studies on SS Cyg \citep{mattei1985}, U Gem  \citep[][]{mayall1957,mattei1987} and VW Hyi \citep{bateson1977}.  To our knowledge, no long-term studies like those of the 1960ies and 1970ies have been published more recently.

Simultaneously, there was and still is a boom of detections of new CVs, especially dwarf novae. The VSX AAVSO catalogue \citep{Watson2006} lists more than 9000 DNe, but does not give any cycle length information. The latest on-line version of the CV catalogue of  \citet[][version 7.24,  February 2017]{ritterykolb} list a total of 852 DNe with known orbital period, 631 of them (74$\%$) are classified as SU UMa, ER UMa or WZ Sge stars.  Only for 121 (19$\%$) some cycle length data are given. Discounting the doubtful cases, \citet{ritterykolb} give only for a total of 73 cases (12$\%$) reliable information on the lengths of the superoutburst cycle (hereafter ``super cycle length").    However, no references are given for the cycle lengths, and it is often very difficult and time consuming to check these data. Therefore, we here determine super cycle length values of other SU UMa stars, for which this knowledge is lacking, or for which only uncertain published information on their superoutburst cadence is available.

\section{ Methods and data sources}

The superoutbursts of most SU UMa stars show a pronounced quasi-periodicity, occurring rather regularly one linear ephemeris with a standard deviation of only 5-10$\%$ of the corresponding period \citep{vogt1980}. This property allows determining a rough value of the cycle length even if only few superoutbursts are recorded. In the present study we limited to those dwarf novae for which no reliable cycle length values are published in the latest on-line version of the catalogue of \citet{ritterykolb}. The main sources were light curves determined by AAVSO members and available in the LCG data base of this organization, as well in the photometric catalogue All Sky Automated Survey ASAS \citep{pojmaski2002}  and in the Catalina Sky Survey CSS \citep{drake2009}. For a few cases this information was complemented from ASAS-SN \citep{Kochanek2017}  and Gaia alerts \citep{Gaiacollaboration2018}. 

The AAVSO archive contains many time-resolved light curves whose details will not be considered in our context. Therefore, we limited our analysis to the V band pass, calculating an average V magnitude value from all observations within the same Julian day.  Since ASAS and CSS contain only observations coinciding accidentally with outburst timings, sometimes it is difficult to decide whether a particular brightening refers to a short eruption or a superoutburst. Criteria are the brightness near maximum, compared to other superoutbursts of the same star, the presence of a continuous bright state during several days, and the expected time differences to previous or following superoutbursts. We include only the epoch (Julian day) of the outburst maximum in our analysis, no other outburst properties as their maximum magnitude, duration, width or decline rate have been determined. Often an observation that was recorded in one of the three above mentioned sources could be confirmed by another one. In each case, a sequence of the cycle count number E was established, beginning with E = 0 for the first recorded superoutburst timing, and assigning the remaining E values based on the minimal time difference between two subsequent records as the basic unit.  A linear least square fit Heliocentric Julian Date (Maximum brightness) vs. E. 

\begin{equation}
\label{eq:ec1}
T_\mathrm{max} = T_\mathrm{0} + C_\mathrm{so} E
\end{equation}

\noindent
leads to the super cycle C$_\mathrm{so}$, its error $\sigma${(C$_\mathrm{so}$)} and the standard deviation which we call here ${\rm \overline{O-C}}$ because it is a measure for the mean difference between the observed superoutburst epoch from the linear ephemeris.

\section{The new catalogue of super cycle length values  }

 Among the SU UMa type dwarf novae considered here there is  only one case  (J1749+1917) that belongs to the ER UMa subclass. A total of 16 stars are of WZ Sge type, most of our remaining targets are ordinary SU UMa stars; five of them could belong to the WZ Sge subclass, as discussed in section 4.3. 
 
 At least three superoutburst timings N have to be available for any star which is included in our analysis. The average number of superoutbursts per target observed is $\overline{N}$ = 6.3, the range is 3 $\leq$ N $\leq$ 18. The results are given in Table~\ref{amplitud}, which lists the 206 dwarf novae with new super cycle values arranged according to right ascension. We use the coordinate identification of \citet{ritterykolb} in cases without official names given in the General Catalogue of Variable Stars (GCVS). Table ~\ref{amplitud} gives all relevant parameters determined according eq.~\ref{eq:ec1} and their errors, as well as the standard deviations $\mathrm{\overline{O-C}}$ and some remarks, mainly referring to WZ Sge classification. Table~\ref{nueva} lists some properties of five stars which also could belong to the WZ Sge class.

\begin{figure}
     \centerline
     \centerline{\includegraphics[width=0.5\textwidth]{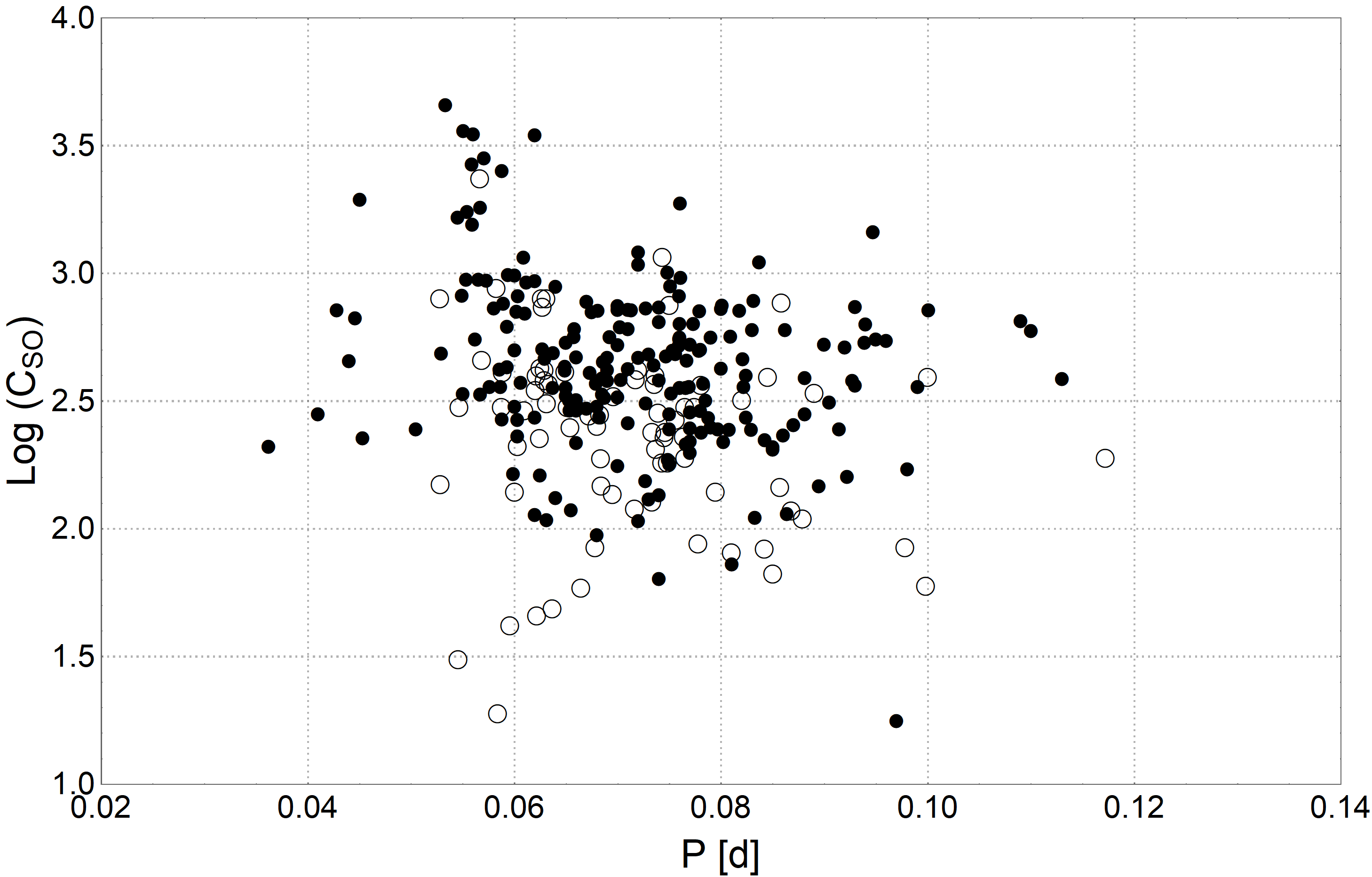}}
      \caption{ Superoutburst cycle length C$_\mathrm{so}$ vs. orbital period P. Filled circles: new C$_\mathrm{so}$ values determined here. Open circles: previously known cycle length values. 
 }
  \label{graLn}
  \end{figure}

We present other two tables as on-line data: Table~\ref{ref} gives additional information on available data used by us for the statistics of Section 4.1 and Fig.~\ref{Cso}  while Table~\ref{Erup} contains all individual superoutburst maximum epochs $T_\mathrm{max}$  and the corresponding cycle numbers E used in our analysis for each of our 206 targets. This table will enable future researchers to combine our results with other(past or future) observations, in order to improve or correct our(presumably in some of our cases)rather preliminary cycle C$_\mathrm{so}$ values.

\section{Results and discussion }

\subsection{$C_{\rm so}$ vs. orbital period and vs. short outburst cycle length }

As mentioned above, only for 73 cases there are reliable superoutburst cycle length values available from the literature. Our study is adding 206 new cases, augmenting the previous knowledge by a factor 2.8. In addition, apart from C$_\mathrm{so}$ and its error, we give here data on the detailed ephemeris of recent superoutbursts. Similar information is available from the analysis of 26 well observed SU UMa stars analyzed by A. Contreras-Quijada in his Master thesis, using a much longer time basis compared to our targets. All these 26 cases form part of the 73 stars listed by  \citet{ritterykolb}, implying that finally only 47 C$_\mathrm{so}$  values are taken from the published literature, and the remaining 232 C$_\mathrm{so}$ values were determined by our group. 

\begin{figure}
     \centering
      \includegraphics[width=0.48\textwidth]{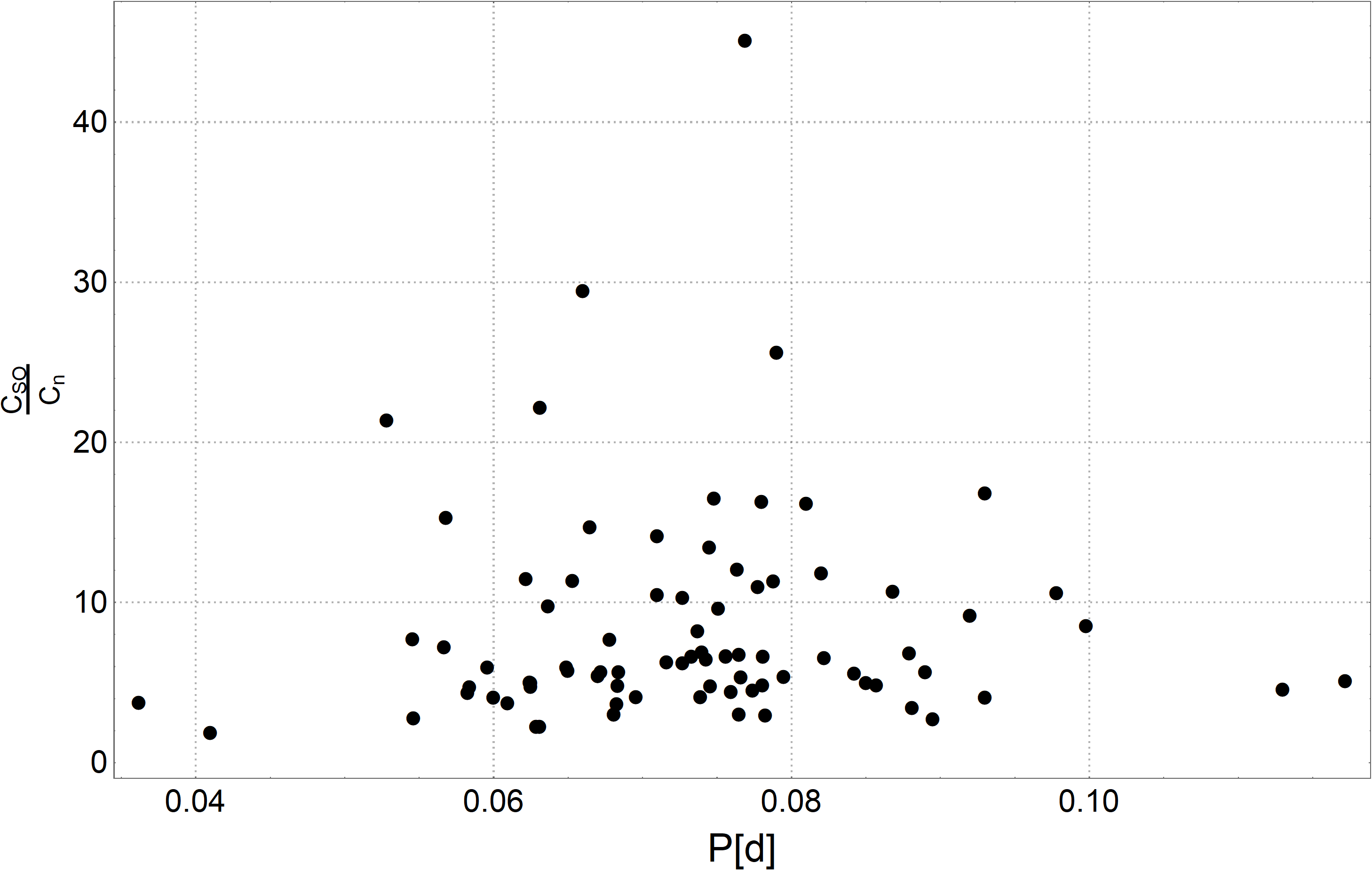}
      \includegraphics[width=0.48\textwidth]{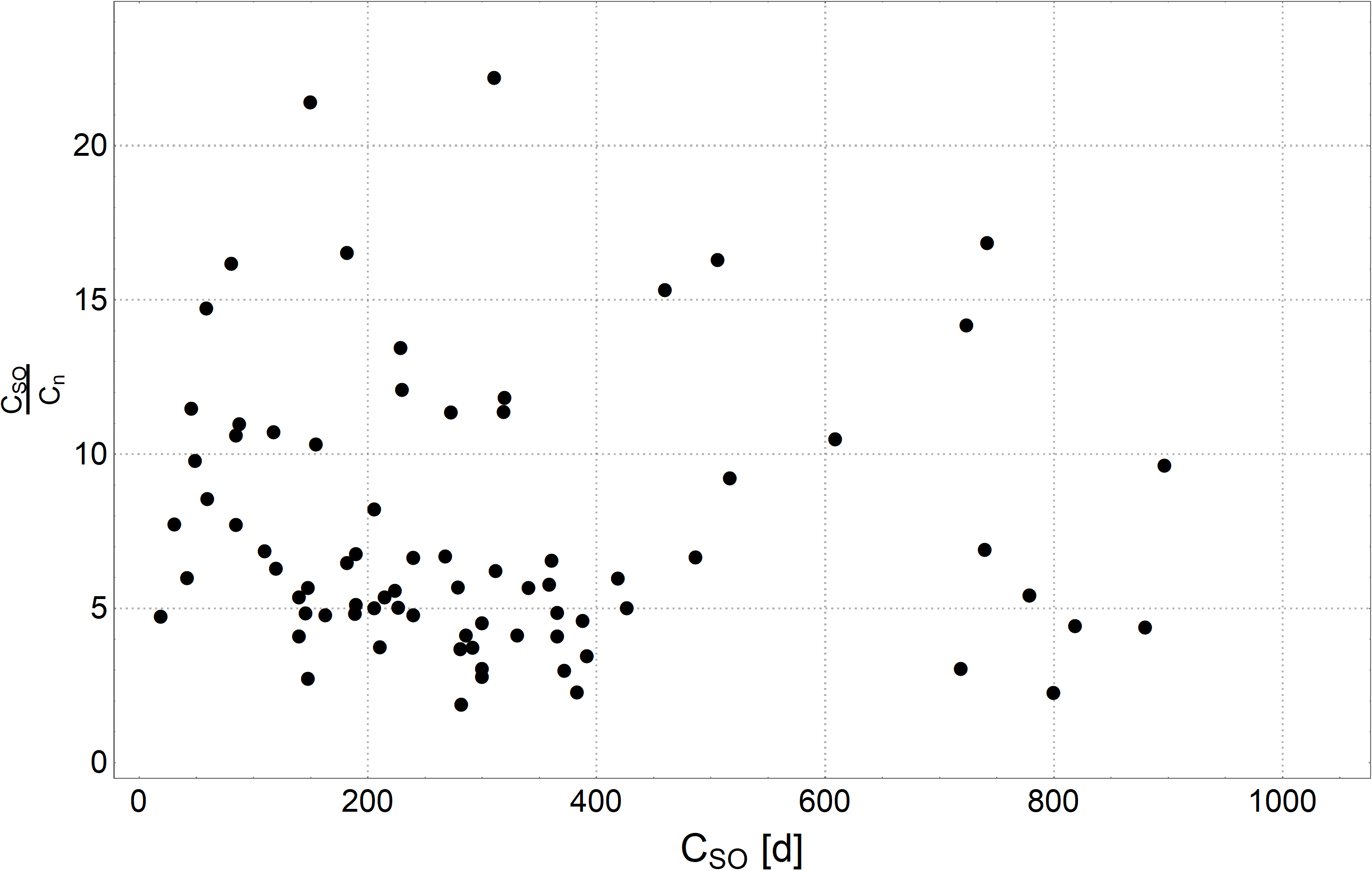}
  \caption{Ratio $C_\mathrm{so}$/$C_\mathrm{n}$ vs. orbital period P (top) and vs. C$_\mathrm{so}$ (bottom).} 
  \label{relacion}
\end{figure}

On the other hand, no attempt was made by us to determine the cycle length C$_\mathrm{n}$ of the ``normal" short outbursts of SU UMa stars. In order to include them in our statistics we use the C$_\mathrm{n}$ values given by \citet{ritterykolb}, complementing them in few cases by cadence data of \citet{coppejans2016}.

\begin{figure}
     \centerline
     \centerline{\includegraphics[width=0.5\textwidth]{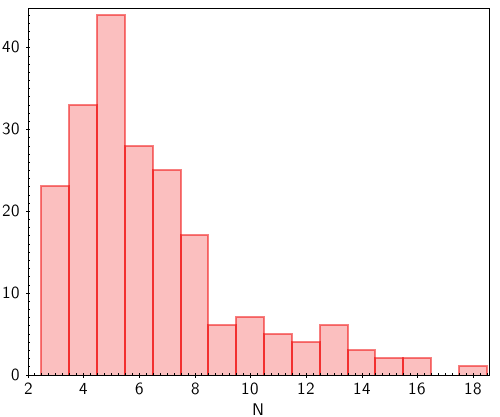}}
      \label{fig:grafT2}
  \caption{ Distribution of the superoutburst numbers N used in the determination of the new cycle length parameters listed in Table~\ref{amplitud}. 
 }
  \label{grafT2}
\end{figure}

Fig.~\ref{graLn} shows the super cycle length values vs. orbital period. There is no clear correlation, but the WZ Sge stars can be distinguished in this diagram in the period range 0.05-0-06 d, due to their large C$_\mathrm{so}$ values. However, there are also ordinary SU UMa stars with shorter super cycle values in this period range. In addition, some of our targets show ultra-short periods (0.035 < Porb < 0.045 d) well separated from the remaining SU UMa stars with Porb > 0.05 d.  They belong to the AM CVn class with orbital periods less than $\sim$ 1 hour, i.e. CVs containing an evolved donor star, typically a Roche lobe filling low mass white dwarf or a semi-degenerate He-rich star \citep{Ramsay2018}.  Despite of their rather different underlying physics (their accretion disks are mainly of helium) their average C$_\mathrm{so}$ values coincide with those of the remaining SU UMa stars with hydrogen rich disks.

In Fig.~\ref{relacion} we show the ratio C$_\mathrm{so}$/C$_\mathrm{n}$ vs. orbital period and vs. C$_\mathrm{so}$.  For C$_\mathrm{so}$ < 400 d seems to be a slight tendency to diminish the frequency of short outbursts for cases with larger superoutburst cycle length, but there is a large scatter in both diagrams.

\begin{figure}
     \centerline
     \centerline{\includegraphics[width=0.5\textwidth]{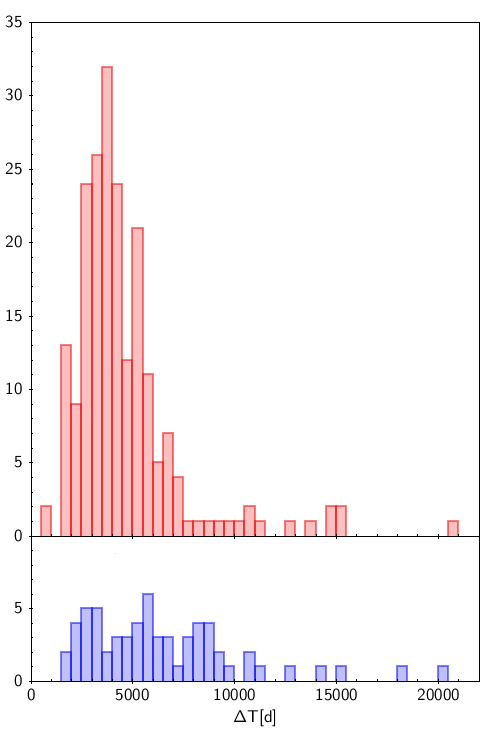}}
      \label{fig:deltaT}
  \caption{  Top: Distribution of the total time intervals $\Delta$T covered by data used for the determination of the new cycle length parameters listed in Table~\ref{amplitud} (binning 500d). Bottom: Distribution of the time intervals of constant C$_\mathrm{so}$ values, found in a study of the long-term cycle length variability of 22 well observed SU UMa stars. Both distributions agree within the expected statistical uncertainties.                 
 }
  \label{deltaT}
\end{figure}

\subsection{Statistical analysis of other parameters   }

In this section we analyze mainly our own data, with emphasis on those parameters that are not available from other published sources. Fig.~\ref{grafT2} shows the distribution of the numbers N of superoutbursts used for our targets, revealing a broad peak around 5 observed superoutburst epochs per star. This could be considered as a rather low number, but, based on the pronounced periodic repeatability of SU UMa type superoutbursts (see below),  it permits a valid first estimation of their cycle length.  Fig.~\ref{deltaT} (upper panel) gives a histogram of the time intervals $\Delta$T between the first and the last observed and catalogued superoutbursts for each star, revealing a maximum around 3000 – 4000 days and a rather symmetric distribution between 1500 and 7000 days, followed by few $\Delta$T values up to more than 20000 days (total range 2 < $\Delta$T < 57 years).  In addition,  Fig.~\ref{deltaT} compares this distribution with results of a recent Master thesis (Contreras-Quijada $\&$ Vogt, in preparation), referring to a study of long-term supercylce length variations of  23 well observed SU UMa stars with mean C$_\mathrm{so}$ values between 88 and 531 days, determined within total time intervals between 19 and 88 years. In 22 cases of this study (96$\%$) significant changes in C$_\mathrm{so}$ in time scales of several years were found, establishing a total of 63 partial time intervals $\Delta$T with constant C$_\mathrm{so}$, whose distribution is shown in the lower panel of Fig.~\ref{deltaT}.  It becomes evident that the distributions in both panels of Fig.~\ref{deltaT} are similar, confirming, that our procedure of linear fits in eq.\eqref{eq:ec1} is valid for a preliminary estimation of C$_\mathrm{so}$, based on our time intervals $\Delta$T. 

This same study permits to estimate the expected accuracy of our C$_\mathrm{so}$ determination. For this purpose, we calculated the individual supercycle lengths values within the time intervals of constant C$_\mathrm{so}$, for each of the above mentioned 22 SU UMa stars with known long-term behavior, and determined the standard deviation $\delta$(C$_\mathrm{so}$)  from the general mean supercycle length ${\rm \overline{C_{So}}}$ in each case. The ratio $\delta$(C$_\mathrm{so}$)/${\rm \overline{C_{So}}}$ is a measure for the reliability of our method, because the time intervals covered by us here correspond to those of typical linear portions in the over-all ${\rm \overline{O-C}}$ vs. E behavior of Contreras-Quijada $\&$ Vogt´s targets, as shown in  Fig.~\ref{deltaT}. The resulting mean value of the ratio $\delta$(C$_\mathrm{so}$)/${\rm \overline{C_{So}}}$ is 0.109 $\pm$ 0.009, with a standard deviation of 0.046, and extremes 0.044 and 0.217. Accordingly, our C$_\mathrm{so}$ values are expected to deviate from those based on long-term studies by $\pm$11$\%$ in average, and should never exceed 22$\%$, implying that our results are valid first approximations.

\begin{figure}
     \centering
     \includegraphics[width=0.5\textwidth]{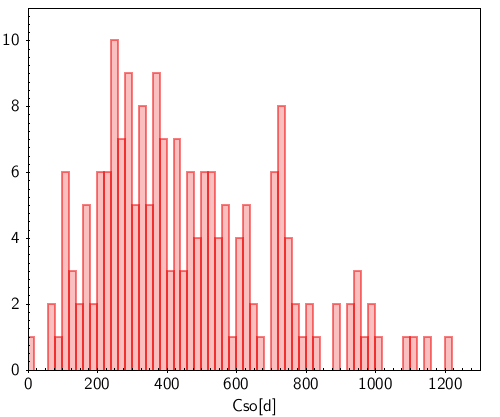}
     \includegraphics[width=0.5\textwidth]{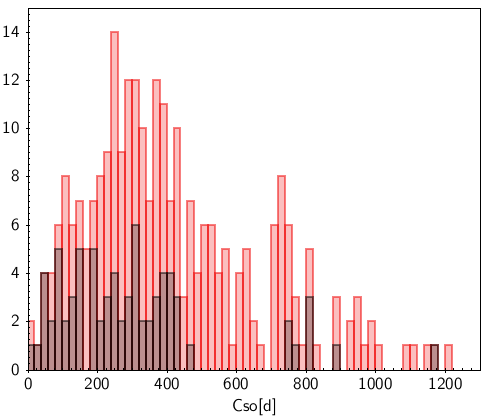}
  \caption{ Top: Distribution of the new C$_\mathrm{so}$ values listed in Table~\ref{amplitud}, limited for C$_\mathrm{so}$ < 1300 d (binnig 20 d). Bottom: this distribution (in red color) is compared and added to that of all previously known  C$_\mathrm{so}$ values (in gris color). It becomes evident that the additional data presented here allow statistical relevant information on the real C$_\mathrm{so}$ distribution for the first time.
  } 
  \label{Cso}
\end{figure}

 The upper panel of  Fig.~\ref{Cso}  shows the C$_\mathrm{so}$ distribution of our targets (with C$_\mathrm{so}$ < 1300 d); in the lower panel of this figure we add to this distribution the hitherto known data, obtaining a statistically significant overall C$_\mathrm{so}$ distribution  for the first time. It reveals  a remarkable maximum around C$_\mathrm{so}$ $\approx$ 270 d which was not present in the earlier data. Generally, there are rapidly increasing numbers for C$_\mathrm{so}$ $\approx$ 100 - 250 d and subsequently slowly declining numbers up to C$_\mathrm{so}$ $\approx$ 800 d.  For larger C$_\mathrm{so}$ values the statistics are still very poor. The narrow maxima around 365 and 730 d could be caused by seasonal selection effects (observational bias of 1 and 2 years). The domain  C$_\mathrm{so}$ > 450 d was unexplored until now; we add a total of 106 cases (51$\%$ of our total sample) to this range of large C$_\mathrm{so}$ values, implying a better statistical basis for future studies of their distribution.

\begin{figure}
     \centering
     \includegraphics[width=0.5\textwidth]{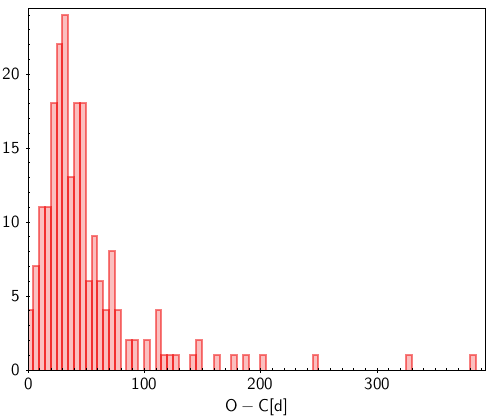}
     \includegraphics[width=0.5\textwidth]{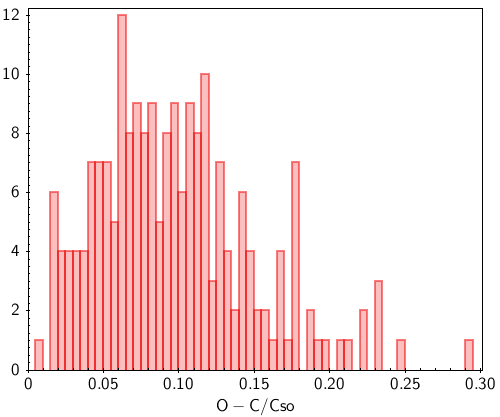}
  \caption{ Distribution of the $\mathrm{\overline{O-C}}$ values (top: binning 5 d ) and of the ratio $\mathrm{\overline{O-C}}$/C$_\mathrm{so}$ of all targets listed in Table~\ref{amplitud} (bottom, binning 0.005d).
  } 
  \label{HistOC}
\end{figure}

Figure~\ref{HistOC} gives histograms of the standard deviations $\mathrm{\overline{O-C}}$ and the ratio  $\mathrm{\overline{O-C}}$/C$_\mathrm{so}$. The $\mathrm{\overline{O-C}}$ distribution is characterized by a broad maximum between 20 and  60 days, with a rapid decline for larger values. The ratio $\mathrm{\overline{O-C}}$/C$_\mathrm{so}$ has a relatively  rapid increase up to a maximum around 0.07 and a slower decline ending at 0.17, with sparse isolated values up to 0.3. This ratio reflects the precision of the ``clock" of quasi-periodic repeatability of superoutbursts, and confirms, that individual superoutburst timings deviate from the overall linear ephemeris in average about $\pm$7$\%$ of the cycle length.  

\begin{figure}
     \centering
     \includegraphics[width=0.5\textwidth]{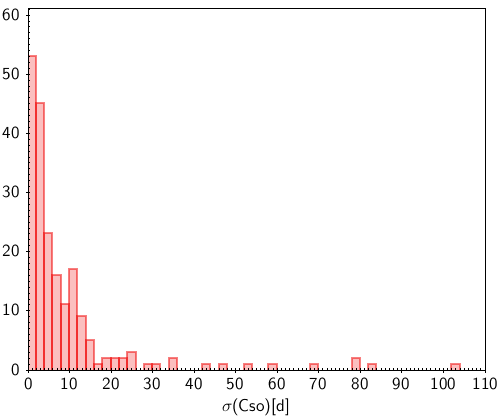}
     \includegraphics[width=0.5\textwidth]{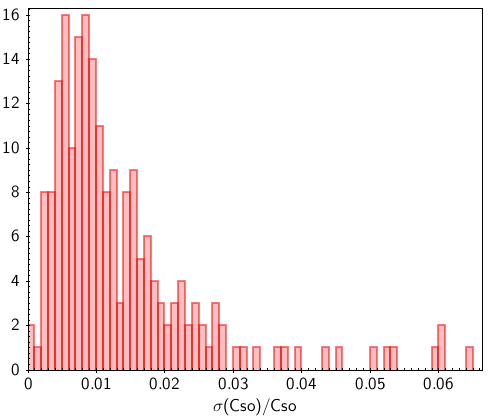}
  \caption{ Distribution of  $\sigma$(C$_\mathrm{so}$) limited to values < 110 d (top: binning 2 d ) and of the ratio $\sigma$(C$_\mathrm{so}$)/C$_\mathrm{so}$ of all targets listed in Table~\ref{amplitud} (bottom, binning 0.001 d.).
  } 
  \label{SigmaCso}
\end{figure}

Figure~\ref{SigmaCso} gives information about the errors $\sigma$(C$_\mathrm{so}$) of the super outburst cycle values. The upper panel shows that these errors are generally very small, mostly < 5 d, with a strong decline until $\sim$ 15 d, and a few outliers with larger values. The lower panel shows the ratio $\sigma$(C$_\mathrm{so}$)/C$_\mathrm{so}$, i.e. the distribution 
of the relative errors of our super cycle length determination and confirm that most   ratios $\sigma$(C$_\mathrm{so}$)/C$_\mathrm{so}$ are rather small with a peak around 0.007 and declining until 0.03, with few larger outlier values $\leq$ 0.06.

\begin{figure}
     \centering
    \includegraphics[width=0.5\textwidth]{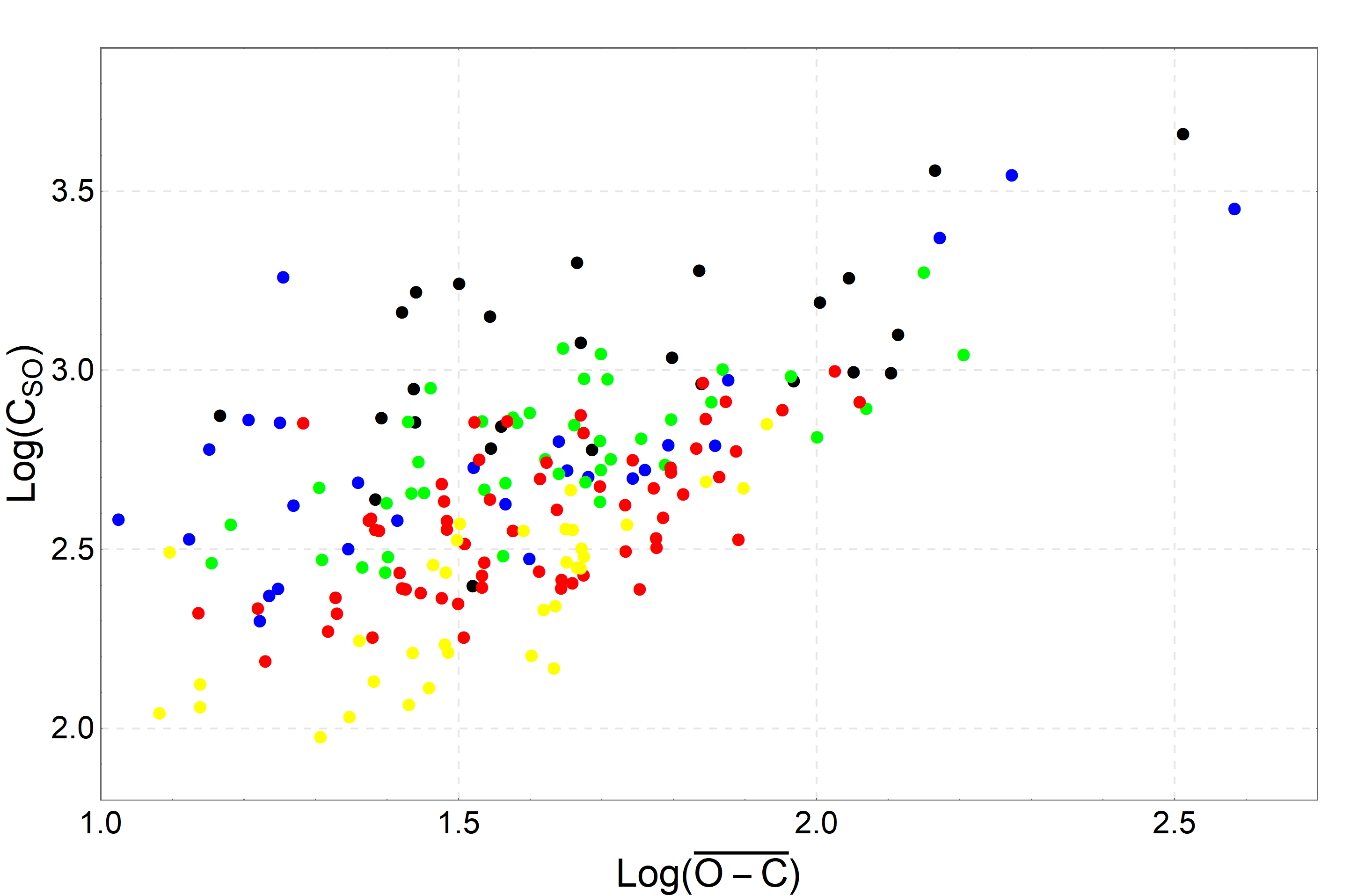}
    \includegraphics[width=0.5\textwidth]{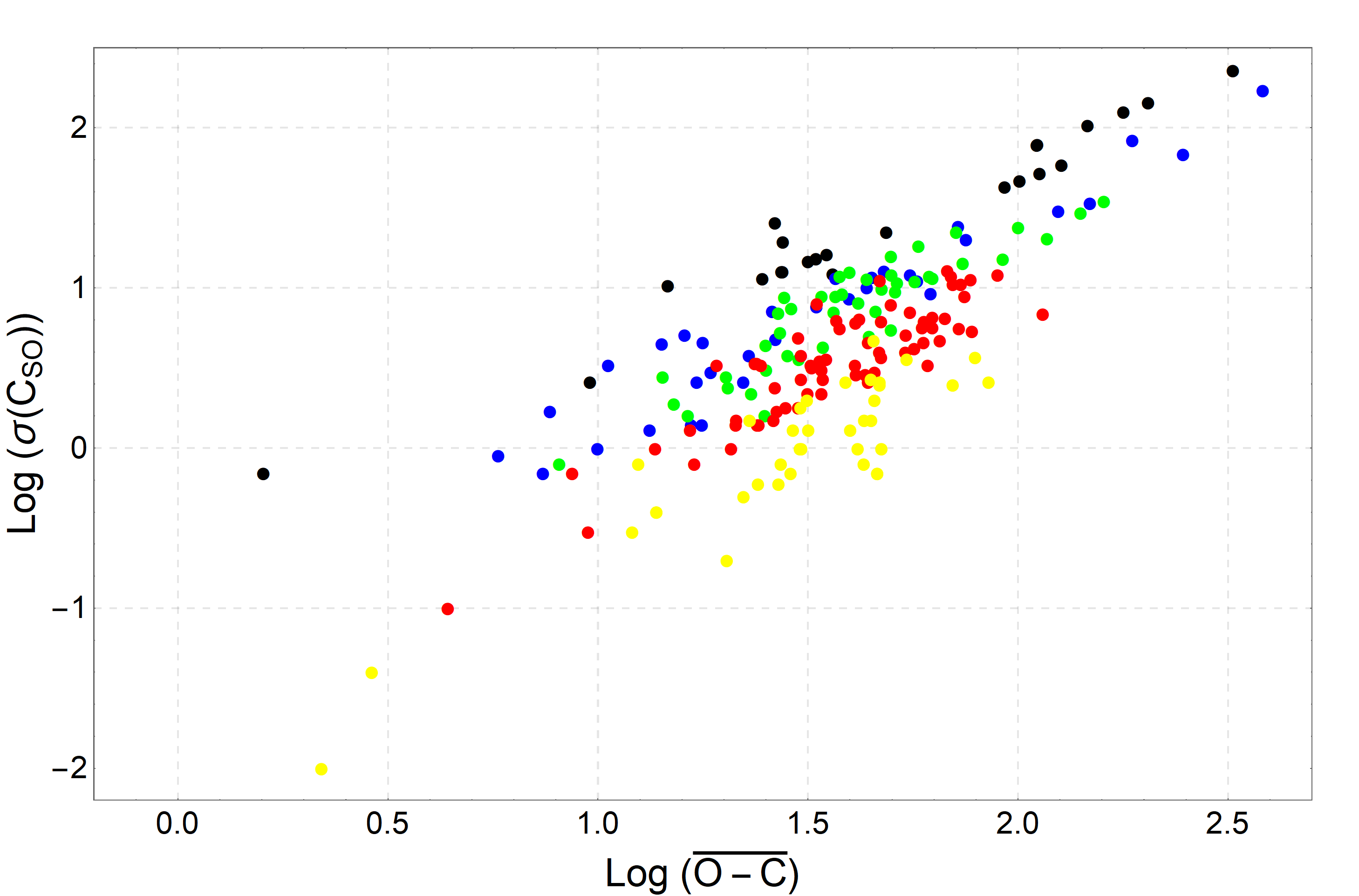}
  \caption{The new super cycle length values  C$_\mathrm{so}$ (top) and their errors  $\sigma$(C$_\mathrm{so}$) (bottom) vs.   $\mathrm{\overline{O-C}}$. Different numbers N are represented by different colors: N=3 (black), N=4 (blue), N=5 (green), N=6-8 (red) and N=9-18 (yellow)
  } 
  \label{logSig}
\end{figure}

Finally, we investigate in Figure~\ref{logSig} how the mean standard deviation $\mathrm{\overline{O-C}}$ is correlated to the super cycle length C$_\mathrm{so}$ and its error $\sigma$(C$_\mathrm{so}$). Four groups of numbers N of observations are indicated by dots of different colors. The upper panel of this figure shows how $\mathrm{\overline{O-C}}$ is correlated to C$_\mathrm{so}$; small numbers N mainly refer to the longest C$_\mathrm{so}$ values, with weak or absent correlation to C$_\mathrm{so}$. In contrast, larger numbers N include also shorter C$_\mathrm{so}$ values, developing systematic trends in the sense that larger cycle lengths are correlated with larger $\mathrm{\overline{O-C}}$ values. According to the lower panel of Figure~\ref{logSig} a general tendency of significant correlations between $\sigma$(C$_\mathrm{so}$) and $\mathrm{\overline{O-C}}$ seems to be valid for all groups of N, now with similar slopes; however these groups are well separated in their zero points: for any given $\mathrm{\overline{O-C}}$ value, group members with N = 3 are characterized by about 10 times larger errors $\sigma$(C$_\mathrm{so}$) than those with 9 $\leq$ N $\leq$ 18, and the groups of  intermediate N values are located between these two extrema. We believe that these features are consequences of the unavoidable inhomogeneity of our data: more observed superoutburst epochs reveal more precise results, underlining the importance to publish always the corresponding error together with any derived parameter.

\textbf{}

\subsection{WZ Sge type stars  }

This  subclass is characterized by long cycle lengths C$_\mathrm{so}$ (between 500 d and up to several decades), large amplitudes in V up to 8$^\mathrm{m}$   and very rare occurrence or absence of short outbursts, revealing only superoutbursts of rather long duration. \citet{KatoT2015} gives some additional properties of WZ Sge stars, as early superhumps, appearing before the ordinary ones, and long or multiple rebrightenings after the main eruption. He discusses many properties of WZ Sge stars in detail, but with little attention to the outburst cycle lengths C$_\mathrm{so}$; he mentions in his Table 6 only outburst years of 58 WZ Sge stars.

 Our targets contain a total of 16 WZ Sge stars, known according to \citet{KatoT2015} and the VSX, (marked “WZ” in Table~\ref{amplitud}) with C$_\mathrm{so}$ values between 2.7 and 12.6 years. On the other hand, we detected five possible WZ Sge candidates among our SU UMa targets, whose most important data are listed in Table~\ref{nueva} (particularly large amplitudes and cycle lengths).  Most of the established WZ Sge stars have short orbital periods between 0.04 and 0.06 days, and none of them > 0.077 days, in contrast to two stars in Table~\ref{nueva} (EF Peg and V444 Peg) that reveal longer orbital periods. Especially V444 Peg could be an interesting case, since it would be the first WZ Sge star within the period gap between 2 and 3 hours. However, all 5 candidates in Table~\ref{nueva} should be confirmed before drawing further conclusions.

 \section{Conclusions }

We have determined individual cycle length values of a total of 206 SU UMa stars whose superoutburst cadence was unknown previously. Using data bases with public access, we were able to increase the number of SU UMa stars with reliable information concerning this crucial parameter by a factor $\sim$ 3.8. To our knowledge, similar studies have not been published for a long time. The dwarf nova cadence statistics by \citet{coppejans2016} cannot be compared to our results because it refers to all types of dwarf novae, and it does not distinguish between short eruptions and superoutbursts  in case of SU UMa stars. 

 Superoutburst cycle length values are now known for a total of 279 SU UMa type dwarf novae. On the other hand, a total of more than 1180 SU UMa stars (including the ER UMa and WZ Sge subtypes) are listed in the AAVSO VSX catalogue, implying still an actual deficit of 76$\%$ when referring to our knowledge of the parameter C$_\mathrm{so}$. On the other hand, we here can present, for the first time, a realistic superoutburst cycle length distribution (Fig.~\ref{Cso}), with a peak around C$_\mathrm{so}$ $\approx$ 270 days. Among our targets there are 16 WZ Sge stars for which we give new C$_\mathrm{so}$ values, and other 5 DNe hitherto classified of SU UMa type, whose long cycles and large amplitudes suggest belonging also to the WZ Sge subgroup. This, however, requires confirmation, especially because two of them seem to exceed the range of orbital periods, typical for WZ Sge stars.

There are striking differences between our study and most similar ones, published recently. In fact, a previous reliable C$_\mathrm{so}$ catalogue does not exist. Sparse information is given by  \citet{ritterykolb}, but without errors and references, and without any indication of possible C$_\mathrm{so}$ variability over years and decades. A first attempt to investigate these items was made by \citet{vogt1980} for few SU UMa stars observed during sufficiently long  time spans at that epoch.  Recently, this was redone by \textbf{Contreras-Quijada $\&$ Vogt} (in preparation) comprising more SU UMa stars and longer time spans now available; this study confirms most of the preliminary results of \citet{vogt1980}.  Based on them, we here compare in  Fig.~\ref{deltaT} our $\Delta$T distribution with that of typical time intervals with constant C$_\mathrm{so}$ cycle lengths determined in well observed SU UMa stars in the course of decades, and find that most SU UMa stars maintain constant C$_\mathrm{so}$ values within similar time intervals as our rather short $\Delta$T values. 
 According to our experience in the above-mentioned study, these “short-lived” C$_\mathrm{so}$ values deviate in average $\pm$11$\%$ from their long-term average in each case. We expect that our 206 new C$_\mathrm{so}$ values comprise uncertainties of this order of magnitude, whenever they will be compared to the (not yet known) long-term behavior of our targets.

Our results are presented in a consciously reproducible way, aimed at enabling any present or future researcher to verify, complement, or modify them, given that they are necessarily limited by the information available today. We present here “early” data, for use in further studies on superoutburst cycles. We regard it as important that there is transparency in the presentation of the data, to avoid the black-box approach of some other studies that renders them difficult to combine with later results.

There is some hope for improvements during coming years. It will be possible to determine many more superoutburst cycle length values whenever future patrol projects will cover suﬃciently large time intervals. In particular, the Vera C. Rubin Observatory \citep[ former LSST, Large Synoptic Survey Telescope,][]{kesler2019} will soon be operational, a crucial step forward in our knowledge of the outburst cadence values of thousands of dwarf novae, including SU UMa stars. Some patience will be necessary, but after several years of operation of the LSST and other patrol projects, the results presented here could be combined with future data, extending this way the total time intervals covered.

\section*{Acknowledgements}
We acknowledge financial support by the Centro de Astrof\'isica de Valpara\'iso (CAV). Part of this work is based on the light curve data base of the American Association of Variable Star Observers (AAVSO) whose rich archival material, mainly based on continuous amateur astronomy efforts, is of great value also for the research by professional astronomers. We thank the referee for constructive comments and advices enabling us to improve the presentation of the paper.  

\section*{Data Availability Statements}

The full data underlying this article are  available as online supplementary material. 

\onecolumn
\begin{table}
\centering
  	\caption{New cycle length values of 206 SU UMa type outbursts. N is the number of superoutbursts covered within the time interval $\Delta$T. The remaining parameters refer to  results  of the linear fit eq.\eqref{eq:ec1}: T$_\mathrm{0}$  is the epoch (in JD-2450000) of the first superoutburst,    $\sigma$(T$_\mathrm{0}$)  its error,  C$_\mathrm{so}$   the supercycle length, $\sigma$(C$_\mathrm{so}$)  its error and  $\mathrm{\overline{O-C}}$ the mean standard deviation of the actual superoutburst epochs from the linear ephemeris eq.\eqref{eq:ec1}. “WZ” refers WZ Sge type stars, “WZ?” to WZ Sge type candidates, also listed in Table~\ref{ref}. The  table is available online. }
	\label{amplitud}
	\begin{tabular}{lcrrrrrrrrr} 
		\hline	                                              
Name &		RA (J2000)&		DEC	&	N	&$\Delta$T  &	T$_\mathrm{0}$	&	$\sigma$(T$_\mathrm{0}$)	&	C$_\mathrm{so}$	&	$\sigma$(C$_\mathrm{so}$)	&	$\mathrm{\overline{O-C}}$& Remarks	 \\
 &	&	&		& (d) &	(d)	&(d)	&(d)	&(d)	& (d)	& \\
    \hline
J0000+3325	&	00 00 24.6	&	+33 25 43	&	6	&	3245	&	4436	&	21	&	247.4	&	2.4	&	26.4	&		\\
BE Oct	&	00 00 48.8	&	-77 18 57	&	6	&	3816	&	4091	&	44	&	476.3	&	7.9	&	49.9	&		\\
J0009-1210	&	00 09 38.2	&	-12 10 16	&	3	&	1822	&	3642	&	47	&	603.0	&	22.5	&	48.6	&		\\
J0015+2636	&	00 15 38.2	&	+26 36 57	&	5	&	3596	&	3689	&	29	&	724.7	&	8.9	&	34.1	&		\\
J0021-5719	&	00 21 30.9	&	-57 19 22	&	5	&	2611	&	2115	&	39	&	529.7	&	12.1	&	50.0	&		\\
J0024-6635	&	00 24 30.7	&	-66 35 53	&	6	&	6303	&	2178	&	48	&	563.7	&	6.5	&	67.1	&		\\
J0036+2151	&	00 36 35.8	&	+21 51 26	&	4	&	3959	&	3329	&	12	&	246.8	&	1.4	&	17.7	&		\\
PT And	&	00 40 24.4	&	+41 04 03	&	4	&	10513	&	-2461	&	157	&	3528.1	&	83.9	&	187.6	&	WZ	\\
LL And	&	00 41 51.5	&	+26 37 21	&	3	&	7274	&	-609	&	134	&	3637.0	&	103.7	&	146.6	&	WZ	\\
J0042-5609	&	00 42 14.2	&	-56 09 20	&	6	&	4254	&	2757	&	20	&	328.9	&	3.2	&	32.3	&		\\
J0045+5032	&	00 45 27.5	&	+50 32 16	&	3	&	762	&	6546	&	32	&	251.0	&	15.3	&	33.1	&		\\
GV Psc	&	01 13 06.7	&	+21 52 51	&	4	&	3702	&	3667	&	46	&	529.9	&	11.1	&	57.6	&		\\
GZ Cet	&	01 37 01.1	&	-09 12 35	&	5	&	5685	&	2050	&	35	&	952.1	&	9.9	&	47.4	&		\\
FZ Cet	&	01 42 25.3	&	-22 15 58	&	5	&	5099	&	2261	&	58	&	733.8	&	11.6	&	62.7	&		\\
J0150+3326	&	01 50 51.5	&	+33 26 22	&	3	&	1837	&	3994	&	34	&	609.2	&	16.3	&	35.1	&		\\
FI Cet	&	01 51 18.6	&	-02 23 01	&	3	&	4645	&	2141	&	85	&	1557.5	&	46.8	&	101.1	&	WZ	\\
BG Ari	&	01 51 51.9	&	+14 00 47	&	4	&	2123	&	4121	&	15	&	235.7	&	2.6	&	17.2	&		\\
J0153+3408	&	01 53 21.5	&	+34 08 56	&	5	&	2936	&	3650	&	29	&	741.7	&	11.9	&	37.7	&		\\
J0211+1716	&	02 11 10.2	&	+17 16 24	&	8	&	5490	&	429	&	16	&	273.2	&	1.5	&	26.2	&		\\
AX For	&	02 19 28.0	&	-30 45 46	&	11	&	5225	&	2225	&	24	&	319.5	&	2.5	&	47.0	&		\\
WY Tri	&	02 25 00.5	&	+32 59 56	&	6	&	5751	&	1805	&	48	&	522.5	&	6.6	&	62.7	&		\\
PU Per	&	02 42 16.1	&	+35 40 46	&	7	&	6207	&	1044	&	58	&	779.3	&	12.1	&	89.7	&		\\
PV Per	&	02 42 53.5	&	+38 04 04	&	6	&	4104	&	12	&	21	&	240.5	&	1.8	&	28.0	&		\\
J0243-1603	&	02 43 54.1	&	-16 03 14	&	3	&	2786	&	3784	&	29	&	700.0	&	12.3	&	36.3	&		\\
BB Ari	&	02 44 57.7	&	+27 31 09	&	4	&	5199	&	3312	&	9	&	399.5	&	1.0	&	10.0	&		\\
UW Tri	&	02 45 17.3	&	+33 31 26	&	3	&	9178	&	-4543	&	297	&	4589.1	&	229.9	&	325.1	&	WZ	\\
RU Hor	&	02 46 07.5	&	-63 35 47	&	6	&	6269	&	86	&	24	&	566.0	&	3.5	&	33.8	&		\\
V368 Per	&	02 47 32.6	&	+34 58 28	&	3	&	3188	&	3752	&	8	&	638.5	&	2.6	&	9.6	&		\\
KY Eri	&	03 10 51.7	&	-07 55 00	&	4	&	1776	&	5127	&	36	&	299.1	&	8.6	&	39.7	&		\\
QY Per	&	03 15 36.8	&	+42 28 14	&	5	&	7693	&	-326	&	71	&	966.8	&	15.2	&	92.2	&		\\
FT Cam	&	03 21 14.4	&	+61 05 27	&	6	&	4480	&	1077	&	29	&	247.3	&	2.6	&	44.0	&		\\
J0334-0710	&	03 34 49.9	&	-07 10 48	&	6	&	5118	&	1968	&	37	&	564.1	&	7.1	&	55.4	&		\\
V701 Tau	&	03 44 01.9	&	+21 57 08	&	8	&	6993	&	43	&	29	&	409.7	&	2.9	&	43.4	&		\\
V1212 Tau	&	03 51 57.0	&	+25 25 28	&	7	&	3633	&	4133	&	27	&	275.9	&	3.3	&	41.0	&		\\
V1024 Per	&	04 02 39.0	&	+42 50 45	&	4	&	1669	&	5625	&	30	&	424.8	&	11.6	&	36.8	&		\\
V1389 Tau	&	04 06 59.8	&	+00 52 44	&	5	&	2564	&	3342	&	39	&	637.6	&	15.8	&	49.9	&		\\
XZ Eri	&	04 11 25.8	&	-15 23 24	&	6	&	6506	&	812	&	55	&	927.7	&	11.9	&	69.4	&		\\
J0411-0907	&	04 11 33.6	&	-09 07 29	&	6	&	2894	&	4109	&	27	&	358.7	&	5.6	&	37.7	&		\\
J0416+2928	&	04 16 36.9	&	+29 28 06	&	7	&	4079	&	3247	&	25	&	291.7	&	2.7	&	34.4	&		\\
V342 Cam	&	04 23 32.9	&	+74 52 50	&	6	&	3997	&	3492	&	31	&	500.3	&	6.1	&	41.1	&		\\
J0426+3541	&	04 26 09.3	&	+35 41 45	&	5	&	1544	&	6209	&	5	&	119.0	&	0.8	&	8.1	&		\\
J0442-0023	&	04 42 16.0	&	-00 23 34	&	4	&	1920	&	3657	&	22	&	383.0	&	7.2	&	26.0	&		\\
HV Aur	&	04 53 16.8	&	+38 16 29	&	10	&	4365	&	3031	&	18	&	274.2	&	1.8	&	30.4	&		\\
V1208 Tau	&	04 59 44.0	&	+19 26 23	&	7	&	4285	&	2790	&	43	&	327.7	&	5.6	&	72.5	&		\\
J0507+1253	&	05 07 16.2	&	+12 53 15	&	13	&	4017	&	3685	&	7	&	133.4	&	0.4	&	13.8	&		\\
GR Ori	&	05 21 35.0	&	+01 10 10	&	3	&	3638	&	2651	&	101	&	1819.2	&	78.5	&	111.0	&	WZ	\\
J0529+1848	&	05 29 58.8	&	+18 48 10	&	13	&	4914	&	2932	&	7	&	95.2	&	0.2	&	20.3	&		\\
V391 Cam	&	05 32 33.9	&	+62 47 52	&	7	&	4381	&	3408	&	26	&	555.4	&	6.4	&	42.0	&		\\
AR Pic	&	05 49 45.4	&	-49 21 56	&	10	&	5466	&	1949	&	23	&	220.5	&	1.5	&	43.2	&		\\
J0557+6832	&	05 57 18.5	&	+68 32 27	&	4	&	3390	&	4074	&	17	&	488.2	&	3.8	&	22.9	&		\\
AD Men	&	06 04 30.9	&	-71 25 23	&	12	&	5152	&	2789	&	23	&	160.6	&	1.3	&	40.0	&		\\
CI Gem	&	06 30 05.9	&	+22 18 51	&	6	&	5392	&	1214	&	54	&	597.5	&	11.3	&	77.3	&		\\
UV Gem	&	06 38 44.1	&	+18 16 11	&	14	&	6817	&	882	&	23	&	147.9	&	0.8	&	43.0	&		\\
PU CMa	&	06 40 47.7	&	-24 23 14	&	11	&	5774	&	1731	&	15	&	337.8	&	2.0	&	31.5	&		\\

\hline  
	\end{tabular}

	\end{table}

\onecolumn
\begin{table}
\centering
  	\contcaption{}
	\begin{tabular}{lcrrrrrrrrr} 
		\hline	                                              
Name &		RA (J2000)&		DEC	&	N	&$\Delta$T  &	T$_\mathrm{0}$	&	$\sigma$(T$_\mathrm{0}$)	&	C$_\mathrm{so}$	&	$\sigma$(C$_\mathrm{so}$)	&	$\mathrm{\overline{O-C}}$& Remarks	 \\
 &	&	&		& (d) &	(d)	&(d)	&(d)	&(d)	& (d)	& \\
    \hline
J0647+4915	&	06 47 25.7	&	+49 15 43	&	6	&	3478	&	4384	&	15	&	217.6	&	1.3	&	16.6	&		\\
CG CMa	&	07 04 05.2	&	-23 45 34	&	3	&	6988	&	1187	&	102	&	3494.0	&	78.8	&	111.4	&	WZ	\\
AQ CMi	&	07 14 34.8	&	+08 48 06	&	7	&	7323	&	113	&	21	&	433.1	&	1.8	&	30.2	&		\\
FQ Mon	&	07 16 41.2	&	-06 56 49	&	5	&	3611	&	3053	&	23	&	721.9	&	7.0	&	26.9	&		\\
V496 Aur	&	07 27 52.3	&	+40 46 54	&	5	&	3327	&	3691	&	21	&	303.0	&	3.1	&	25.2	&		\\
J0732+4130	&	07 32 08.1	&	+41 30 09	&	6	&	3236	&	3809	&	24	&	248.9	&	3.1	&	34.1	&		\\
J0741-0945	&	07 41 12.7	&	-09 45 56	&	5	&	3982	&	3135	&	19	&	283.1	&	2.2	&	23.2	&		\\
J0746+1734	&	07 46 40.6	&	+17 34 13	&	4	&	3201	&	3835	&	17	&	318.9	&	2.6	&	22.2	&		\\
J0800+1924	&	08 00 33.9	&	+19 24 16	&	6	&	3977	&	3460	&	58	&	506.1	&	10.6	&	73.2	&		\\
J0803+2516	&	08 03 03.9	&	+25 16 27	&	7	&	2897	&	3463	&	29	&	260.9	&	4.6	&	44.1	&		\\
J0803+2848	&	08 03 07.0	&	+28 48 56	&	7	&	3707	&	3708	&	11	&	154.6	&	0.8	&	17.0	&		\\
KK Cnc	&	08 07 14.3	&	+11 38 13	&	3	&	2995	&	4483	&	107	&	987.0	&	58.8	&	127.1	&	WZ	\\
LX Cnc	&	08 12 07.6	&	+13 18 24	&	5	&	2391	&	3652	&	28	&	486.9	&	8.9	&	36.8	&		\\
J0813-0103	&	08 13 18.5	&	-01 03 28	&	6	&	5056	&	2677	&	24	&	268.5	&	2.2	&	34.1	&		\\
J0814+0907	&	08 14 08.4	&	+09 07 59	&	5	&	3323	&	4050	&	47	&	547.1	&	11.9	&	61.5	&		\\
J0814-0050	&	08 14 18.9	&	-00 50 22	&	6	&	4685	&	2713	&	13	&	187.4	&	1.0	&	20.8	&		\\
J0819+1915	&	08 19 36.1	&	+19 15 40	&	3	&	2214	&	4883	&	24	&	740.1	&	11.5	&	24.7	&		\\
FL Lyn	&	08 24 09.7	&	+49 31 24	&	7	&	3167	&	4113	&	41	&	321.2	&	6.2	&	59.8	&		\\
J0838+4910	&	08 38 45.2	&	+49 10 56	&	4	&	1545	&	4893	&	9	&	385.2	&	3.3	&	10.6	&		\\
RX Vol	&	08 39 32.3	&	-66 17 39	&	8	&	4002	&	3089	&	18	&	232.4	&	1.8	&	30.0	&		\\
J0841+2100	&	08 41 27.4	&	+21 00 54	&	8	&	3130	&	4352	&	13	&	210.4	&	1.5	&	21.4	&		\\
EG Cnc	&	08 43 04.0	&	+27 51 50	&	4	&	12617	&	-4395	&	211	&	2529.8	&	68.6	&	247.3	&	WZ	\\
V498 Hya	&	08 45 55.1	&	+03 39 30	&	3	&	2947	&	4521	&	109	&	992.3	&	52.2	&	112.8	&		\\
CT Hya	&	08 51 07.4	&	+03 08 34	&	7	&	4977	&	4039	&	31	&	419.2	&	4.0	&	46.8	&		\\
J0851+3444	&	08 51 13.4	&	+34 44 49	&	4	&	2912	&	4783	&	13	&	731.6	&	5.1	&	16.1	&		\\
BZ UMa	&	08 53 44.1	&	+57 48 41	&	9	&	15198	&	-8904	&	33	&	303.4	&	1.0	&	47.4	&		\\
PU UMa	&	09 01 03.9	&	+48 09 11	&	4	&	3037	&	4426	&	41	&	501.3	&	12.1	&	55.4	&		\\
MM Hya	&	09 14 14.0	&	-06 47 45	&	9	&	5498	&	1310	&	24	&	362.3	&	2.7	&	44.7	&		\\
GZ Cnc	&	09 15 51.7	&	+09 00 50	&	9	&	5064	&	2120	&	31	&	281.9	&	2.6	&	46.9	&		\\
J0943-2720	&	09 43 27.3	&	-27 20 30	&	7	&	5170	&	1928	&	36	&	470.6	&	5.7	&	59.3	&		\\
J0945-1944	&	09 45 51.0	&	-19 44 01	&	5	&	3972	&	1987	&	32	&	567.4	&	8.1	&	41.8	&		\\
J0947+0610	&	09 47 59.8	&	+06 10 44	&	3	&	5367	&	3577	&	186	&	2682.9	&	144.4	&	204.2	&	WZ	\\
J0959-1601	&	09 59 26.5	&	-16 01 47	&	7	&	4414	&	1877	&	16	&	232.9	&	1.4	&	21.3	&		\\
J1005+1911	&	10 05 15.4	&	+19 11 08	&	5	&	4482	&	2994	&	21	&	897.1	&	7.5	&	28.9	&		\\
CP Dra	&	10 15 39.8	&	+73 26 05	&	8	&	4655	&	1947	&	16	&	245.9	&	1.7	&	26.7	&		\\
NSV 4838	&	10 23 20.2	&	+44 05 09	&	5	&	3718	&	3774	&	10	&	372.8	&	1.9	&	15.2	&		\\
J1026+1920	&	10 26 16.1	&	+19 20 45	&	5	&	2973	&	3817	&	19	&	427.3	&	4.4	&	25.1	&		\\
J1027-4343	&	10 27 05.8	&	-43 43 41	&	5	&	1756	&	4184	&	11	&	291.0	&	2.8	&	14.3	&		\\
J1028-0819	&	10 28 42.9	&	-08 19 27	&	8	&	3143	&	2796	&	9	&	211.1	&	1.0	&	13.7	&		\\
RX Cha	&	10 36 26.3	&	-80 02 48	&	5	&	3575	&	1296	&	29	&	717.9	&	9.2	&	38.2	&		\\
J1100+1315	&	11 00 14.7	&	+13 15 52	&	5	&	3795	&	3372	&	15	&	472.0	&	2.8	&	20.2	&		\\
J1100-1156	&	11 00 38.1	&	-11 56 47	&	4	&	4402	&	2735	&	9	&	339.0	&	1.3	&	13.3	&		\\
TU Crt	&	11 03 36.5	&	-21 37 46	&	9	&	4697	&	2058	&	26	&	465.1	&	4.7	&	45.4	&		\\
J1122-1110	&	11 22 53.3	&	-11 10 38	&	3	&	685	&	4666	&	1	&	228.2	&	0.7	&	1.6	&		\\
RZ Leo	&	11 37 22.3	&	+01 48 58	&	5	&	11353	&	-3796	&	104	&	1888.1	&	29.6	&	141.3	&	WZ	\\
V359 Cen	&	11 58 15.3	&	-41 46 08	&	7	&	5030	&	1707	&	13	&	716.0	&	3.3	&	19.2	&		\\
J1200-1526	&	12 00 52.9	&	-15 26 21	&	5	&	4413	&	1957	&	14	&	366.2	&	1.6	&	16.4	&		\\
J1202+4503	&	12 02 31.0	&	+45 03 49	&	5	&	5203	&	1880	&	16	&	274.0	&	1.6	&	25.0	&		\\
GP CVn	&	12 27 40.8	&	+51 39 25	&	5	&	4227	&	3754	&	24	&	467.4	&	4.3	&	34.4	&		\\
AL Com	&	12 32 25.9	&	+14 20 43	&	4	&	13542	&	-7419	&	134	&	2358.9	&	34.0	&	148.7	&	WZ	\\
V591 Cen	&	12 42 18.0	&	-33 34 07	&	5	&	4338	&	1417	&	36	&	431.8	&	5.5	&	49.9	&		\\
GO Com	&	12 56 37.1	&	+26 36 44	&	7	&	3608	&	2842	&	46	&	607.9	&	12.9	&	68.0	&		\\
V485 Cen	&	12 57 23.3	&	-33 12 07	&	18	&	14976	&	-7724	&	21	&	282.4	&	0.7	&	46.4	&		\\
HV Vir	&	13 21 03.2	&	+01 53 29	&	4	&	6193	&	-1015	&	321	&	2840.5	&	171.6	&	383.6	&	WZ	\\
LY Hya	&	13 31 53.8	&	-29 40 59	&	5	&	5964	&	870	&	56	&	1010.5	&	14.3	&	74.0	&	WZ?	\\

\hline  
	\end{tabular}

	\end{table}

\onecolumn
\begin{table}
\centering
  	\contcaption{}
	\begin{tabular}{lcrrrrrrrrr}  
		\hline	                                              
Name &		RA (J2000)&		DEC	&	N	&$\Delta$T  &	T$_\mathrm{0}$	&	$\sigma$(T$_\mathrm{0}$)	&	C$_\mathrm{so}$	&	$\sigma$(C$_\mathrm{so}$)	&	$\mathrm{\overline{O-C}}$& Remarks	 \\
 &	&	&		& (d) &	(d)	&(d)	&(d)	&(d)	& (d)	& \\
    \hline

J1343-4426	&	13 43 37.2	&	-44 26 42	&	10	&	5082	&	2044	&	15	&	130.5	&	0.7	&	28.8	&		\\
HS Vir	&	13 43 38.4	&	-08 14 03	&	14	&	6545	&	973	&	22	&	361.0	&	2.0	&	45.7	&		\\
J1425+1515	&	14 25 48.1	&	+15 15 01	&	4	&	3806	&	3795	&	39	&	635.6	&	10.1	&	43.7	&		\\
OU Vir	&	14 35 00.1	&	-00 46 07	&	6	&	5910	&	1613	&	57	&	735.2	&	10.6	&	70.1	&		\\
V362 Lib	&	14 43 41.9	&	-17 55 49	&	4	&	2489	&	2410	&	66	&	619.5	&	24.4	&	72.2	&		\\
UZ Boo	&	14 44 01.3	&	+22 00 56	&	5	&	12725	&	-6225	&	40	&	1157.8	&	5.0	&	44.3	&	WZ	\\
J1459+3548	&	14 59 21.8	&	+35 48 05	&	5	&	1844	&	4560	&	28	&	305.0	&	7.1	&	36.5	&		\\
NZ Boo	&	15 02 41.0	&	+33 34 24	&	5	&	3093	&	4203	&	31	&	763.7	&	12.6	&	39.8	&		\\
V418 Ser	&	15 14 53.6	&	+02 09 35	&	5	&	2734	&	4500	&	21	&	456.4	&	5.3	&	27.2	&		\\
DM Dra	&	15 34 12.1	&	+59 48 32	&	7	&	4352	&	1817	&	22	&	438.7	&	3.6	&	35.0	&		\\
AB Nor	&	15 49 15.5	&	-43 04 49	&	4	&	1772	&	752	&	4	&	221.2	&	0.9	&	5.8	&		\\
V493 Ser	&	15 56 44.3	&	-00 09 51	&	6	&	3758	&	3815	&	34	&	752.8	&	11.2	&	46.9	&		\\
QZ Ser	&	15 56 54.5	&	+21 07 20	&	5	&	5385	&	858	&	90	&	784.9	&	20.4	&	117.4	&		\\
VW CrB	&	16 00 03.8	&	+33 11 15	&	7	&	5085	&	2073	&	24	&	723.6	&	6.3	&	37.0	&		\\
V893 Sco	&	16 15 15.1	&	-28 37 31	&	5	&	2243	&	4313	&	22	&	557.8	&	8.8	&	27.8	&		\\
V589 Her	&	16 22 07.2	&	+19 22 36	&	8	&	3479	&	1430	&	36	&	314.3	&	5.1	&	54.2	&		\\
V699 Oph	&	16 25 14.8	&	-04 40 26	&	8	&	5610	&	1219	&	59	&	337.9	&	5.4	&	77.9	&		\\
J1625+1203	&	16 25 20.3	&	+12 03 09	&	4	&	1728	&	5398	&	20	&	247.0	&	4.8	&	26.5	&		\\
V1032 Oph	&	16 26 09.7	&	-03 53 25	&	7	&	2691	&	4921	&	2.7	&	72.6	&	0.1	&	4.4	&		\\
J1628+0653	&	16 28 06.2	&	+06 53 15	&	5	&	2947	&	3617	&	13	&	296.9	&	2.4	&	20.4	&		\\
J1628+2402	&	16 28 30.9	&	+24 02 59	&	7	&	3064	&	3454	&	15	&	382.3	&	3.4	&	23.7	&		\\
FL TrA	&	16 30 36.6	&	-61 50 21	&	8	&	3563	&	3571	&	31	&	269.6	&	3.7	&	47.3	&		\\
J1631+1031	&	16 31 20.9	&	+10 31 34	&	4	&	2555	&	4570	&	45	&	506.6	&	12.8	&	48.0	&		\\
J1639+1224	&	16 39 42.7	&	+12 24 14	&	6	&	2870	&	3565	&	6	&	205.7	&	0.7	&	8.7	&		\\
V611 Her	&	16 44 49.4	&	+19 59 40	&	7	&	3436	&	3678	&	7	&	114.4	&	0.3	&	9.5	&		\\
J1649+0358	&	16 49 50.4	&	+03 58 35	&	5	&	3944	&	3235	&	40	&	648.6	&	11.0	&	56.9	&		\\
DV Sco	&	16 50 27.9	&	-28 07 59	&	5	&	2191	&	2514	&	45	&	555.3	&	18.4	&	58.1	&		\\
V1227 Her	&	16 53 59.1	&	+20 10 10	&	16	&	4098	&	3593	&	14	&	117.1	&	0.6	&	27.0	&		\\
V1239 Her	&	17 02 13.3	&	+32 29 54	&	6	&	3595	&	3655	&	24	&	720.0	&	8.0	&	33.3	&		\\
V877 Ara	&	17 16 53.9	&	-65 32 52	&	5	&	3432	&	2491	&	38	&	567.5	&	10.8	&	51.6	&		\\
V2527 Oph	&	17 22 04.4	&	-19 49 09	&	3	&	1502	&	3202	&	13	&	751.1	&	10.4	&	14.7	&		\\
J1730+6247	&	17 30 08.3	&	+62 47 55	&	11	&	5341	&	2143	&	23	&	358.5	&	2.6	&	39.0	&		\\
MM Sco	&	17 30 45.3	&	-42 11 41	&	15	&	20625	&	-13188	&	42	&	711.6	&	2.6	&	85.3	&		\\
J1740+4147	&	17 40 33.5	&	+41 47 56	&	3	&	3900	&	4308	&	163	&	1949.9	&	126.2	&	178.4	&	WZ	\\
V660 Her	&	17 42 09.2	&	+23 48 31	&	11	&	5209	&	1333	&	28	&	372.3	&	3.6	&	54.4	&		\\
J1743+2311	&	17 43 05.7	&	+23 11 08	&	4	&	4342	&	4232	&	5	&	334.0	&	0.7	&	7.4	&		\\
J1748+5050	&	17 48 27.9	&	+50 50 40	&	4	&	2417	&	4747	&	11	&	603.9	&	4.5	&	14.2	&		\\
J1749+1913	&	17 49 02.1	&	+19 13 30	&	13	&	3649	&	2836	&	1	&	17.8	&	0.01	&	2.2	&		\\
NY Her	&	17 52 52.6	&	+29 22 19	&	13	&	4402	&	3559	&	2	&	63.7	&	0.04	&	2.9	&		\\
J1821+1709	&	18 21 13.9	&	+17 09 16	&	7	&	1975	&	2759	&	23	&	180.2	&	3.3	&	32.2	&		\\
J1830+1233	&	18 30 01.8	&	+12 33 48	&	10	&	4995	&	2814	&	16	&	172.3	&	1.0	&	30.3	&		\\
DT Oct	&	18 40 52.6	&	-83 43 10	&	10	&	4369	&	2642	&	7	&	311.6	&	0.8	&	12.5	&		\\
J1900-4930	&	19 00 05.3	&	-49 30 34	&	6	&	5722	&	2885	&	35	&	358.8	&	2.9	&	41.2	&		\\
V419 Lyr	&	19 10 13.9	&	+29 06 14	&	14	&	3684	&	2890	&	7	&	115.2	&	0.4	&	13.8	&		\\
V585 Lyr	&	19 13 58.4	&	+40 44 09	&	4	&	4966	&	2250	&	47	&	622.3	&	9.3	&	62.1	&		\\
V344 Pav	&	19 16 09.7	&	-62 35 53	&	10	&	6058	&	1468	&	15	&	287.7	&	1.3	&	29.2	&		\\
J1927-4847	&	19 27 10.1	&	-48 47 52	&	10	&	2974	&	2437	&	14	&	176.6	&	1.5	&	23.0	&		\\
V2176 Cyg	&	19 27 11.6	&	+54 17 52	&	3	&	5252	&	672	&	27	&	1753.4	&	14.7	&	31.7	&	WZ	\\
KX Aql	&	19 33 53.7	&	+14 17 47	&	6	&	14826	&	-7859	&	85	&	819.2	&	6.9	&	114.9	&		\\
V1006 Cyg	&	19 48 47.2	&	+57 09 22	&	8	&	4008	&	3958	&	20	&	361.6	&	2.7	&	30.5	&		\\
V1047 Aql	&	19 51 31.1	&	+10 57 22	&	12	&	4815	&	3092	&	13	&	108.2	&	0.5	&	22.3	&		\\
DO Vul	&	19 52 10.7	&	+19 34 43	&	4	&	4673	&	2804	&	55	&	944.2	&	20.1	&	75.3	&		\\
V405 Vul	&	19 53 05.0	&	+21 14 49	&	8	&	3130	&	2773	&	15	&	388.0	&	3.4	&	23.9	&		\\
V1454 Cyg	&	19 53 38.5	&	+35 21 46	&	5	&	6647	&	333	&	47	&	949.5	&	9.5	&	51.1	&		\\
V725 Aql	&	19 56 45.1	&	+10 49 32	&	7	&	5904	&	-174	&	37	&	537.8	&	5.7	&	62.6	&		\\

\hline  
	\end{tabular}

	\end{table}

\onecolumn
\begin{table}
\centering
  	\contcaption{}
	\begin{tabular}{lcrrrrrrrrr} 
		\hline	                                              
Name &		RA (J2000)&		DEC	&	N	&$\Delta$T  &	T$_\mathrm{0}$	&	$\sigma$(T$_\mathrm{0}$)	&	C$_\mathrm{so}$	&	$\sigma$(C$_\mathrm{so}$)	&	$\mathrm{\overline{O-C}}$& Remarks	 \\
 &	&	&		& (d) &	(d)	&(d)	&(d)	&(d)	& (d)	& \\
    \hline
AW Sge	&	19 58 37.1	&	+16 41 29	&	9	&	9927	&	-3372	&	52	&	471.1	&	3.7	&	79.2	&		\\
V337 Cyg	&	19 59 52.9	&	+39 14 00	&	8	&	7460	&	299	&	41	&	389.7	&	3.3	&	61.1	&		\\
V1028 Cyg	&	20 00 56.6	&	+56 56 37	&	12	&	10145	&	-4165	&	22	&	375.4	&	1.3	&	31.8	&		\\
V550 Cyg	&	20 05 05.2	&	+32 21 21	&	5	&	5652	&	-327	&	34	&	707.9	&	7.2	&	45.9	&		\\
AX Cap	&	20 08 57.0	&	-17 16 38	&	5	&	3365	&	3221	&	77	&	654.7	&	24	&	100.3	&		\\
J2009-6326	&	20 09 24.1	&	-63 26 23	&	5	&	5081	&	2987	&	26	&	504.8	&	3.6	&	30.1	&	WZ?	\\
V1316 Cyg	&	20 12 13.6	&	+42 45 52	&	12	&	5568	&	2118	&	12	&	136.2	&	0.6	&	24.1	&		\\
J2027-2240	&	20 27 31.2	&	-22 40 01	&	4	&	2634	&	2020	&	35	&	527.7	&	11.7	&	44.9	&		\\
SY Cap	&	20 29 47.5	&	-15 54 38	&	13	&	4879	&	2037	&	17	&	163.2	&	0.8	&	27.3	&		\\
HO Del	&	20 36 55.5	&	+14 0310	&	8	&	7411	&	-391	&	48	&	823.2	&	8.9	&	74.7	&		\\
IL Vul	&	20 38 32.7	&	+22 42 17	&	6	&	3360	&	3554	&	21	&	484.4	&	4.9	&	30.0	&		\\
J2054-1940	&	20 54 08.2	&	-19 40 27	&	5	&	2584	&	3993	&	37	&	517.3	&	11.4	&	43.7	&		\\
QU Aqr	&	21 00 14.1	&	+00 44 46	&	16	&	4023	&	3589	&	5	&	111.0	&	0.3	&	12.1	&		\\
J2102+0258	&	21 02 05.7	&	+02 58 34	&	7	&	3955	&	3693	&	18	&	180.4	&	1.4	&	24.0	&		\\
AO Oct	&	21 05 08.2	&	-75 21 02	&	13	&	8733	&	-1416	&	26	&	292.5	&	1.5	&	44.8	&		\\
V364 Peg	&	21 12 29.7	&	+12 32 04	&	7	&	6512	&	798	&	16	&	361.0	&	1.4	&	24.2	&		\\
EF Peg	&	21 15 04.1	&	+14 03 49	&	5	&	6640	&	-1385	&	110	&	1110.7	&	34.9	&	160.5	&	WZ?	\\
J2120+1941	&	21 20 25.2	&	+19 41 57	&	3	&	2838	&	4048	&	79	&	937.7	&	43.1	&	93.0	&		\\
J2123-1539	&	21 23 25.6	&	-15 39 54	&	4	&	6067	&	2252	&	98	&	1214.9	&	30.3	&	124.8	&	WZ?	\\
J2125-1026	&	21 25 21.8	&	-10 26 28	&	4	&	3213	&	3714	&	13	&	200.4	&	1.4	&	16.7	&		\\
V630 Cyg	&	21 34 59.2	&	+40 40 19	&	15	&	8021	&	-3613	&	20	&	215.3	&	1.0	&	41.6	&		\\
V632 Cyg	&	21 36 04.2	&	+40 26 19	&	11	&	15345	&	-9491	&	53	&	491.4	&	2.5	&	70.2	&		\\
V444 Peg	&	21 37 01.8	&	+07 14 46	&	3	&	2923	&	4760	&	33	&	1461.4	&	25.7	&	26.4	&	WZ?	\\
V627 Peg	&	21 38 06.7	&	+26 19 57	&	3	&	3321	&	3643	&	25	&	1660.6	&	19.5	&	27.6	&	WZ	\\
J2144+2220	&	21 44 26.4	&	+22 20 25	&	3	&	2667	&	3532	&	26	&	891.4	&	12.7	&	27.4	&		\\
J2147+2445	&	21 47 38.4	&	+24 45 54	&	6	&	3850	&	3530	&	28	&	382.1	&	3.8	&	30.5	&		\\
V476 Peg	&	21 54 33.7	&	+35 50 17	&	8	&	3958	&	3683	&	21	&	358.8	&	3.3	&	24.5	&		\\
J2158+0947	&	21 58 15.3	&	+09 47 10	&	6	&	4865	&	2909	&	42	&	341.1	&	4.6	&	59.7	&		\\
J2206+3014	&	22 06 41.1	&	+30 14 36	&	4	&	3376	&	3653	&	17	&	421.1	&	3.0	&	18.6	&		\\
J2212+1601	&	22 12 32.0	&	+16 01 40	&	7	&	4623	&	2899	&	33	&	255.7	&	3.0	&	45.6	&		\\
V521 Peg	&	22 21 44.8	&	+18 40 08	&	9	&	10591	&	2881	&	17	&	164.1	&	1.0	&	30.6	&		\\
J2234-0355	&	22 34 18.4	&	-03 55 30	&	4	&	2355	&	3525	&	7	&	391.8	&	1.7	&	7.7	&		\\
V650 Peg	&	22 43 48.5	&	+08 09 27	&	4	&	3609	&	3664	&	16	&	719.0	&	4.6	&	17.8	&		\\
TY PsA	&	22 49 39.9	&	-27 06 54	&	8	&	3172	&	1883	&	18	&	224.4	&	2.2	&	31.6	&		\\
J2258-0949	&	22 58 31.2	&	-09 49 32	&	6	&	3730	&	2874	&	35	&	246.0	&	4.2	&	56.7	&		\\
V368 Peg	&	22 58 43.5	&	+11 09 12	&	8	&	6269	&	1401	&	39	&	452.5	&	4.7	&	65.2	&		\\
HY Psc	&	23 03 51.7	&	+01 06 52	&	5	&	4128	&	2480	&	19	&	457.5	&	3.8	&	28.3	&		\\
CC Scl	&	23 15 31.9	&	-30 48 46	&	8	&	5960	&	1733	&	32	&	423.4	&	4.0	&	54.1	&		\\
V776 And	&	23 19 36.1	&	+36 47 00	&	4	&	3234	&	3692	&	30	&	538.6	&	7.7	&	33.2	&		\\
EG Aqr	&	23 25 19.2	&	-08 18 19	&	5	&	3210	&	4044	&	55	&	818.5	&	22.5	&	71.3	&		\\
EI Psc	&	23 29 54.3	&	+06 28 11	&	7	&	5421	&	2257	&	31	&	671.7	&	6.2	&	47.3	&		\\
J2333-1557	&	23 33 13.0	&	-15 57 44	&	3	&	2152	&	4392	&	27	&	719.8	&	12.7	&	27.5	&		\\

\hline  
	\end{tabular}

	\end{table}

\newpage

\twocolumn

\begin{table}
\centering
  	\caption{New WZ Sge type candidate stars. 	P$_\mathrm{o}$   and   P$_\mathrm{s}$ refer to orbital and superhump periods resp.,  A to the outburst amplitude in V. N and C$_\mathrm{so}$  as in Table~\ref{amplitud}.  References:(1) \citet{Still1994}, (2) \citet{Howell1993},\\
  (3) \citet{KatoT2014}. 
       }
	\label{nueva}
		\scalebox{0.77} {
	\begin{tabular}{lcrrcrrr} 
		\hline	         
Name &	Other name&		P$_\mathrm{o}$	& P$_\mathrm{s}$ & A & N& C$_\mathrm{so}$& Ref \\
   &		&		(d)	& (d) & (mag) & &(d)& \\
	\hline
LY Hya & 1329-294& 0.0748 &  0.077  & 5.3 & 5 & 1011& (1) \\
J2009-6326&  ASASSN-14lk   &   0.0600 &   0.0614 &    7.5     &    5      &    505  &-\\
EF Peg   & AN 143.1935 &       0.0837  &   0.087  &    7.8 &  5 &   1111&       (2) \\                          
J2123-1539  &   ASASSN-14dh &     0.072  &    0.0736  &   7.3  &       4   & 1215  &      -\\
V444 Peg & Var Peg 2008  &  0.0947 &    0.0995&   6.4 &  3 &1461  &     (3) \\

\hline 
  	\end{tabular}
  	}

	\end{table}

\begin{table}
\centering
  	\caption{Data used for our statistical comparison with published information. C$_\mathrm{n}$ refer to    cycle length values of normal (short) outbursts of SU UMa stars, C$_\mathrm{so}$ to cycle lengths of their superoutbursts. The references in the last column are: (1) this work, only for C$_\mathrm{so}$; (2) Contreras-Quijada et al, in preparation, only for C$_\mathrm{so}$; (3) \citet{coppejans2016} only for C$_\mathrm{n}$; (4) \citet[][version 7.24,  February 2017]{ritterykolb}    for C$_\mathrm{n}$ and/or C$_\mathrm{so}$ in the remaining cases. The full table is available online.
  }
	\label{ref}
	\begin{tabular}{lcrrrr} 
		\hline	         
Name &		RA (J2000)&		DEC	& C$_\mathrm{n}$ & C$_\mathrm{so}$ & Ref \\
   &		&			& (d) & (d) &  \\
	\hline
J0000+3325	&	00 00 24.6	&	+33 25 43	&	-	&	247	&	1	\\
BE Oct	&	00 00 48.8	&	-77 18 57	&	-	&	476	&	1	\\
J0009-1210	&	00 09 38.2	&	-12 10 16	&	-	&	603	&	1	\\
V402 And	&	00 11 07.3	&	+30 32 36	&	-	&	400	&	4	\\
J0015+2636	&	00 15 38.2	&	+26 36 57	&	-	&	725	&	1	\\
J0021-5719	&	00 21 30.9	&	-57 19 22	&	-	&	530	&	1	\\

\hline  
	\end{tabular}

	\end{table}

\begin{table}
\centering
  	\caption{Cycle count numbers E and epochs T$_\mathrm{max}$ (JD-2450000) of  individual superoutbursts used for the calculation of the parameters in Table~\ref{amplitud}. The full table is available online.  }
	\label{Erup}
	\begin{tabular}{lcrr} 
		\hline	         
Name &		E &		T$_\mathrm{max}$ & \\
   &		& (d)  \\
	\hline
J0000+3325	&	0	&	4417.5&	\\
	&	4	&	5449.4&	\\
	&	6	&	5944.6&	\\
	&	9	&	6628.4&	\\
	&	12	&	7399.6&	\\
	&	13	&	7662.56	\\
	\hline 
BE Oct	&	0	&	4123&	\\
	&	4	&	5919&	\\
	&	5	&	6482&	\\
	&	6	&	6975.1&	\\
	&	7	&	7397.1&	\\
	&	8	&	7938.8&	\\
	\hline 
J0009-1210	&	0	&	3654.5&	\\
	&	2	&	4808.5&	\\
	&	3	&	5476.5&	\\
	\hline 
J0015+2636	&	0	&	3709.5&	\\
	&	2	&	5092.4	&\\
	&	3	&	5866.5	&\\
	&	4	&	6617.6	&\\
	&	5	&	7305.4	&\\
	\hline 
J0021-5719	&	0	&	2168.7&	\\
	&	1	&	2578.7	&\\
	&	3	&	3703.7	&\\
	&	4	&	4230.9	&\\
	&	5	&	4779.6	&\\

\hline  
	\end{tabular}

	\end{table}

\newpage






\bibliographystyle{mnras}
\bibliography{referencia}

\begin{thebibliography}{}
\makeatletter
\relax
\def\mn@urlcharsother{\let\do\@makeother \do\$\do\&\do\#\do\^\do\_\do\%\do\~}
\def\mn@doi{\begingroup\mn@urlcharsother \@ifnextchar [ {\mn@doi@}
  {\mn@doi@[]}}
\def\mn@doi@[#1]#2{\def\@tempa{#1}\ifx\@tempa\@empty \href
  {http://dx.doi.org/#2} {doi:#2}\else \href {http://dx.doi.org/#2} {#1}\fi
  \endgroup}
\def\mn@eprint#1#2{\mn@eprint@#1:#2::\@nil}
\def\mn@eprint@arXiv#1{\href {http://arxiv.org/abs/#1} {{\tt arXiv:#1}}}
\def\mn@eprint@dblp#1{\href {http://dblp.uni-trier.de/rec/bibtex/#1.xml}
  {dblp:#1}}
\def\mn@eprint@#1:#2:#3:#4\@nil{\def\@tempa {#1}\def\@tempb {#2}\def\@tempc
  {#3}\ifx \@tempc \@empty \let \@tempc \@tempb \let \@tempb \@tempa \fi \ifx
  \@tempb \@empty \def\@tempb {arXiv}\fi \@ifundefined
  {mn@eprint@\@tempb}{\@tempb:\@tempc}{\expandafter \expandafter \csname
  mn@eprint@\@tempb\endcsname \expandafter{\@tempc}}}

\bibitem[\protect\citeauthoryear{{Bateson}}{{Bateson}}{1977}]{bateson1977}
{Bateson} F.~M.,  1977, New Zealand Journal of Science, \href
  {https://ui.adsabs.harvard.edu/abs/1977NZJS...20...73B} {20, 73}

\bibitem[\protect\citeauthoryear{{Coppejans}, {K{\"o}rding}, {Knigge},
  {Pretorius}, {Woudt}, {Groot}, {Van Eck}  \& {Drake}}{{Coppejans}
  et~al.}{2016}]{coppejans2016}
{Coppejans} D.~L.,  {K{\"o}rding} E.~G.,  {Knigge} C.,  {Pretorius} M.~L.,
  {Woudt} P.~A.,  {Groot} P.~J.,  {Van Eck} C.~L.,   {Drake} A.~J.,  2016,
  \mn@doi [\mnras] {10.1093/mnras/stv2921}, \href
  {https://ui.adsabs.harvard.edu/abs/2016MNRAS.456.4441C} {456, 4441}

\bibitem[\protect\citeauthoryear{{Drake} et~al.,}{{Drake}
  et~al.}{2009}]{drake2009}
{Drake} A.~J.,  et~al., 2009, \mn@doi [\apj] {10.1088/0004-637X/696/1/870},
  \href {https://ui.adsabs.harvard.edu/abs/2009ApJ...696..870D} {696, 870}

\bibitem[\protect\citeauthoryear{{Gaia Collaboration} et~al.,}{{Gaia
  Collaboration} et~al.}{2018}]{Gaiacollaboration2018}
{Gaia Collaboration} et~al., 2018, \mn@doi [\aap]
  {10.1051/0004-6361/201833051}, \href
  {https://ui.adsabs.harvard.edu/abs/2018A&A...616A...1G} {616, A1}

\bibitem[\protect\citeauthoryear{{Howell}, {Schmidt}, {De Young}, {Fried},
  {Schmeer}  \& {Gritz}}{{Howell} et~al.}{1993}]{Howell1993}
{Howell} S.~B.,  {Schmidt} R.,  {De Young} J.~A.,  {Fried} R.,  {Schmeer} P.,
  {Gritz} L.,  1993, \mn@doi [\pasp] {10.1086/133198}, \href
  {https://ui.adsabs.harvard.edu/abs/1993PASP..105..579H} {105, 579}

\bibitem[\protect\citeauthoryear{{Kato}}{{Kato}}{2015}]{KatoT2015}
{Kato} T.,  2015, \mn@doi [\pasj] {10.1093/pasj/psv077}, \href
  {https://ui.adsabs.harvard.edu/abs/2015PASJ...67..108K} {67, 108}

\bibitem[\protect\citeauthoryear{{Kato} et~al.,}{{Kato}
  et~al.}{2014}]{KatoT2014}
{Kato} T.,  et~al., 2014, \mn@doi [\pasj] {10.1093/pasj/psu072}, \href
  {https://ui.adsabs.harvard.edu/abs/2014PASJ...66...90K} {66, 90}

\bibitem[\protect\citeauthoryear{{Kato} et~al.,}{{Kato} et~al.}{2017}]{kato1}
{Kato} T.,  et~al., 2017, \mn@doi [\pasj] {10.1093/pasj/psx058}, \href
  {https://ui.adsabs.harvard.edu/abs/2017PASJ...69...75K} {69, 75}

\bibitem[\protect\citeauthoryear{{Kessler} et~al.,}{{Kessler}
  et~al.}{2019}]{kesler2019}
{Kessler} R.,  et~al., 2019, \mn@doi [\pasp] {10.1088/1538-3873/ab26f1}, \href
  {https://ui.adsabs.harvard.edu/abs/2019PASP..131i4501K} {131, 094501}

\bibitem[\protect\citeauthoryear{{Kochanek} et~al.,}{{Kochanek}
  et~al.}{2017}]{Kochanek2017}
{Kochanek} C.~S.,  et~al., 2017, \mn@doi [\pasp] {10.1088/1538-3873/aa80d9},
  \href {https://ui.adsabs.harvard.edu/abs/2017PASP..129j4502K} {129, 104502}

\bibitem[\protect\citeauthoryear{{Mattei}, {Saladyga}  \& {Waagen}}{{Mattei}
  et~al.}{1985}]{mattei1985}
{Mattei} J.~A.,  {Saladyga} M.,   {Waagen} E.~O.,  1985, {SS Cygni light curves
  1896-1985}

\bibitem[\protect\citeauthoryear{{Mattei}, {Saladyga}, {Waagen}  \&
  {Jones}}{{Mattei} et~al.}{1987}]{mattei1987}
{Mattei} J.~A.,  {Saladyga} M.,  {Waagen} E.~O.,   {Jones} C.~M.,  1987, {U
  Geminorum light curves 1855-1985}

\bibitem[\protect\citeauthoryear{{Mayall}}{{Mayall}}{1957}]{mayall1957}
{Mayall} M.~W.,  1957, \jrasc, \href
  {https://ui.adsabs.harvard.edu/abs/1957JRASC..51..165M} {51, 165}

\bibitem[\protect\citeauthoryear{{Patterson}}{{Patterson}}{1999}]{patterson1}
{Patterson} J.,  1999, in {Mineshige} S.,  {Wheeler} J.~C.,  eds,  Vol. 26,
  Disk Instabilities in Close Binary Systems. p.~61

\bibitem[\protect\citeauthoryear{{Pojmanski}}{{Pojmanski}}{2002}]{pojmaski2002}
{Pojmanski} G.,  2002, \actaa, \href
  {https://ui.adsabs.harvard.edu/abs/2002AcA....52..397P} {52, 397}

\bibitem[\protect\citeauthoryear{{Ramsay} et~al.,}{{Ramsay}
  et~al.}{2018}]{Ramsay2018}
{Ramsay} G.,  et~al., 2018, \mn@doi [\aap] {10.1051/0004-6361/201834261}, \href
  {https://ui.adsabs.harvard.edu/abs/2018A&A...620A.141R} {620, A141}

\bibitem[\protect\citeauthoryear{{Ritter} \& {Kolb}}{{Ritter} \&
  {Kolb}}{2003}]{ritterykolb}
{Ritter} H.,  {Kolb} U.,  2003, \mn@doi [\aap] {10.1051/0004-6361:20030330},
  \href {https://ui.adsabs.harvard.edu/abs/2003A%26A...404..301R} {404,
  301(version 7.24, February 2017)}

\bibitem[\protect\citeauthoryear{{Still}, {Marsh}, {Dhillon}  \&
  {Horne}}{{Still} et~al.}{1994}]{Still1994}
{Still} M.~D.,  {Marsh} T.~R.,  {Dhillon} V.~S.,   {Horne} K.,  1994, \mn@doi
  [\mnras] {10.1093/mnras/267.4.957}, \href
  {https://ui.adsabs.harvard.edu/abs/1994MNRAS.267..957S} {267, 957}

\bibitem[\protect\citeauthoryear{{Vogt}}{{Vogt}}{1974}]{vogt1}
{Vogt} N.,  1974, \aap, \href
  {https://ui.adsabs.harvard.edu/abs/1974A%26A....36..369V} {36, 369}

\bibitem[\protect\citeauthoryear{{Vogt}}{{Vogt}}{1980}]{vogt1980}
{Vogt} N.,  1980, \aap, \href
  {https://ui.adsabs.harvard.edu/abs/1980A%26A....88...66V} {88, 66}

\bibitem[\protect\citeauthoryear{{Warner}}{{Warner}}{1975}]{warner2}
{Warner} B.,  1975, \mn@doi [\mnras] {10.1093/mnras/170.1.219}, \href
  {https://ui.adsabs.harvard.edu/abs/1975MNRAS.170..219W} {170, 219}

\bibitem[\protect\citeauthoryear{{Warner}}{{Warner}}{1995}]{warner9}
{Warner} B.,  1995, Cambridge Astrophysics Series, \href
  {https://ui.adsabs.harvard.edu/abs/1995CAS....28.....W} {28}

\bibitem[\protect\citeauthoryear{{Watson}}{{Watson}}{2006}]{Watson2006}
{Watson} C.~L.,  2006, Journal of the American Association of Variable Star
  Observers (JAAVSO), \href
  {https://ui.adsabs.harvard.edu/abs/2006JAVSO..35..318W} {35, 318}

\makeatother
\end{thebibliography}




\bsp	
\label{lastpage}
\end{document}